\documentclass[aps,pre,reprint,groupedaddress,showkeys]{revtex4-1}

\usepackage{graphicx,subfigure}%
\usepackage{amsmath,amssymb}
\usepackage{bm}

\begin{document}


\title{Fluctuating micro-heterogeneity in water--tert-butyl alcohol mixtures and lambda-type divergence of the mean cluster size with phase transition-like multiple anomalies}



\author{Saikat Banerjee}
\author{Jonathan Furtado}
\author{Biman Bagchi}
\email{bbagchi@sscu.iisc.ernet.in}
\affiliation{Solid State and Structural Chemistry Unit, Indian Institute of Science, Bangalore 560012, India}


\date{\today}

\begin{abstract}
Water--tert-butyl alcohol (TBA) binary mixture exhibits a large number of thermodynamic and dynamic anomalies. 
These anomalies are observed at surprisingly low TBA mole fraction, with  $x_{\text{\tiny{TBA}}} \approx 0.03 - 0.07$. 
We demonstrate here that the origin of the anomalies lies in the local structural changes that occur due to 
self-aggregation of TBA molecules. We observe a percolation transition of the TBA molecules 
at $x_{\text{\tiny{TBA}}} \approx 0.05$. We note that ``islands'' of TBA clusters form even below this mole 
fraction, while a large spanning cluster emerges above that mole fraction. At this percolation threshold, 
we observe a lambda-type divergence in the fluctuation of the size of the largest TBA cluster, 
reminiscent of a critical point. Alongside, the structure of water is also perturbed, albeit weakly, 
by the aggregation of TBA molecules. There is a monotonic decrease in the tetrahedral order parameter of water, 
while the dipole moment correlation shows a weak non-linearity. 
Interestingly, water molecules themselves exhibit a reverse percolation transition at higher TBA concentration, 
$x_{\text{\tiny{TBA}}} \approx 0.45$, where large spanning water clusters now break-up into small clusters. 
This is accompanied by significant divergence of the fluctuations in the size of largest water cluster. 
This second transition gives rise to another set of anomalies around. Both the percolation transitions can be 
regarded as manifestations of Janus effect at small molecular level.
\end{abstract}


\maketitle

\section{\label{introduction} Introduction}
Ever since the seminal review of Franks and Ives~\cite{franks_ives_qrev_1966} in 1966, 
where they endorsed the idea of structural transformations as the underlying cause for many 
anomalies observed in water-alcohol systems, the microscopic origin of the anomalous behavior 
of amphiphilic solutes in water has been frequently discussed. Examples of such non-ideal behavior
in the binary mixture of water and amphiphilic solutes are varied and well-studied
~\cite{franks_ives_qrev_1966, frank_evans_jcp_1945,
woolf_jced_2009, koga_pccp_1999, nakanishi_jpc_1967, nakanishi_jpc_1970, kaatze_jpca_2000, kaatze_jpca_1999,
tomasik_jpoc_1990, ranjit_jpca_2008, ranjit_jchemsci_2008, ranjit_jpca_2011, jolicoeur_jsolchem_1982,
seifer_jstructchem_1978, easteal_bcsj_1994, desnoyers_canjc_1977}.
Sometimes the anomaly is strongest at surprisingly low solute concentrations.

Dramatic anomalies in thermodynamic and dynamic properties are routinely observed in aqueous binary 
mixtures of many amphiphilic solutes like methanol, dimethyl sulfoxide(DMSO), ethanol, dioxane, phenol, 
glycerol, etc. Unfortunately however these systems are studied and reported as individual cases, without 
addressing the scope of a general unifying understanding that could have far reaching consequence. 
We have earlier reported studies of water--DMSO~\cite{saikat_jpcb_2010_1, saikat_jpcb_2010_2} and 
water-ethanol~\cite{saikat_jpcb_2012} mixtures. The striking similarity of water-TBA binary mixture to 
water-DMSO and water-ethanol solutions certainly points to the above mentioned possibility of a unified 
understanding of the anomalies in all these binary mixtures, in terms of hydrophobic-hydrophilic character 
of these amphiphilic molecules.

As the amphiphilic solutes contain both hydrophobic and hydrophilic groups, the same molecule can 
induce opposite effects in water. While the hydrophilic groups can interact favorably with water 
(forming strong H-bonds), the hydrophobic groups tend to self-aggregate and disrupt the water structure 
by hydrophobic hydration. Such dual effects are often referred to as ``Janus Effect''~\cite{granick_science_2002} 
after the name of the Greek God Janus, with two faces one facing forward and the other opposite. 
These two opposing effects combine together to modify the extensive H-bond network of water in their 
aqueous binary mixtures~\cite{lum_chandler_weeks, chandler_nature_2005}. 
The idea of structural orientation of water molecules surrounding an alcohol molecule was first conceived by 
Frank and Evans~\cite{frank_evans_jcp_1945}. They suggested that the non-polar residues of the alcohol 
molecules reinforce low entropy water caging, with strong H-bonds in the first hydration shell of the alcohols. 
This creates an open network structure of water (as in low temperature water, or ice). The solute then goes into 
the open network structure of bulk water, thus reducing the total volume required. This picture, 
popularly known as the ``iceberg'' model, received broad support from different studies in the years to follow.

However, this concept fell short in explaining the modern diffraction experiments~\cite{soper_finney_prl_1993,
soper_finney_nature_2002, soper_finney_epl_2002, soper_finney_jcp_2004, finney_bowron_biophyschem_2003}. 
The alternative picture proposed self-association of the amphiphilic solutes to form hydrophobic 
aggregates~\cite{chandler_nature_2005, soper_finney_jpcb_2005}. The hydrophobic aggregates lead to a 
microheterogeneity in the system, though it remains homogeneous in the macroscopic scale. 
The properties of the aggregates and the physical nature of microheterogeneity are not fully understood, 
and remain a subject of current research and debate.

Among monohydric alcohols that are miscible with water at any proportions, tertiary butyl alcohol (TBA) possesses 
the largest aliphatic group. The hydrophobic interaction is, therefore, much higher than many other amphiphilic 
small molecules. This is manifested as strong anomalies in many physical and thermodynamic properties. 
Partial molal volume of TBA, as evaluated from density measurements by Nakanishi~\cite{nakanishi_bcsj_1960}, 
showed a sharp minimum at  $x_{\text{\tiny{TBA}}} \approx 0.03$. Visser et al used flow microcalorimetric 
technique to calculate the heat capacity of water-TBA mixture~\cite{desnoyers_canjc_1977}. 
The heat capacity increases up to $x_{\text{\tiny{TBA}}} \approx 0.05$ and then falls to its molar value -- 
a trend which is similar to the aqueous solution of surfactants. Hence, they suggested some kind of microphase 
transition, similar to micellization. Iwasaki and Fujiyama~\cite{fujiyama_jpc_1979} predicted encaged TBA 
molecules surrounded by H-bonded water molecules (the then-popular ``iceberg'' model) following their Rayleigh 
light scattering experiment, which showed that concentration fluctuation of water-TBA mixture, beyond 
$x_{\text{\tiny{TBA}}} \approx 0.05$, deviated abruptly from ideal values with increase in temperature.
Bender and Pecora~\cite{bender_pecora_jpc_1986} detected considerable dispersion in ultrasonic speed of 
sound in water-TBA mixture over a concentration range 0.0-0.16 during Brillouin scattering experiment. 
Structural relaxation of water molecules surrounding the solute was used to rationalize the observations. 
In yet another light scattering experiment, Vuks and Shurupova~\cite{shurupova_optcommun_1972} found an additional 
maximum at $x_{\text{\tiny{TBA}}} \approx 0.03$ (apart from the theoretically predicted one) and interpreted 
the same using phase transition. However, their results were later contended~\cite{jolly_optcommun_1974}.

Recently, Egorov and Makarov~\cite{makarov_jchemthermo_2011} did an extensive study on volume and density 
properties of water-TBA mixture over the whole concentration range. They found anomalous behavior in 
excess molar volume, thermal isobaric expansivity, partial molar volume and partial thermal isobaric expansivity. 
Despite the obvious consensus on the thermodynamics of the association process, the molecular level 
structure has been highly debated. Apart from the general success of ``iceberg'' model, another explanation 
involved the clathrate-hydrate structure, wherein the TBA molecules were supposed to form clathrates with 
water molecules. Such clustering phenomenon was used to explain many experimental results. As for example, 
Euliss and Sorensen~\cite{sorensen_jcp_1984} interpreted the anomaly in correlation length of concentration 
fluctuation, measured by photon correlation spectroscopy (PCS), in terms of TBA and water clathrate aggregates. 
Similarly, the microscopic concentration fluctuations observed in static and dynamic light scattering 
experiment by Subramanian et al~\cite{sengers_jced_2011} was speculated to be arising from metastable 
clathrate-like precursors triggered by minute traces of impurities. However, such explanations were questioned 
by the diffraction experiments. Bowron, Finney and Soper~\cite{soper_finney_bowron_jpcb_1998, 
soper_finney_bowron_jcp_2001} observed ``dominant non-polar to non-polar solute contacts'' in water-TBA 
binary mixture, particularly at mole fraction 0.06, in their neutron diffraction study using 
hydrogen / deuterium isotopic substitution. The aggregation of TBA molecules close to the anomalous 
concentration range was further supported by direct structural evidence from small angle neutron 
scattering experiments~\cite{soper_finney_jpcb_2005, teixeira_jcsfaraday_trans_1990}. Contrary to the 
earlier speculation of enhancement / destruction of water structure in an alcohol solution, these experiments 
showed that the local structure of water is surprisingly close to that of bulk water. 
The excess entropy arises from incomplete mixing at the molecular level, rather than from water restructuring. 
Concentration fluctuations, obtained from small angle X-ray scattering measurements by Nishikawa et 
al.~\cite{nishikawa_jpc_1987}, though originally interpreted using the clathrate-like structure of water 
in the first hydration shell, actually corroborates to the idea of microsegregation.

In their specially designed mass spectrometric studies, Wakisaka and co-workers~\cite{wakisaka_jpca_2004} 
found direct evidence of self-aggregation of TBA. Such microsegregation, also sometimes referred as 
microheterogeneity, is observed even in aqueous methanol -- the lowest member of the homologous series of 
monohydric alcohols. Study of absorption spectra using chromophores also showed the existence of the 
aggregates~\cite{ranjit_jpca_2008, ranjit_jchemsci_2008, ranjit_jpca_2011}.

Given the controversy over the exact molecular picture in water-TBA mixture, one might naturally expect 
that computer simulation studies would be able to sort out the scenario. Indeed, several simulation 
studies~\cite{nakanishi_fluidpheq_1993, kusalik_jpcb_2000_watertba, fornili_pccp_2003, kerdcharoen_cpl_2003,
paul_patey_jpcb_2006, lee_vegt_tba_pot, perera_jcp_2012, gupta_patey_jcp_2012, slipchenko_jpcb_2012} 
have been performed on this important binary mixture, at different concentrations and using different force fields. 
Early in the 1990's, Tanaka and Nakanishi~\cite{nakanishi_fluidpheq_1993} performed simulations on this system and 
noted self-aggregation of TBA molecules at mole fraction 0.17. However, no such clustering was observed at 
$x_{\text{\tiny{TBA}}} = 0.08$ or $0.03$.  Later in 2000, Kusalik et al focused on the structure of the binary mixture 
at $x_{\text{\tiny{TBA}}} = 0.02$ and $0.08$ via molecular dynamics simulation with two different TBA force 
fields~\cite{kusalik_jpcb_2000_watertba}. Contrary to the earlier results, they observed spontaneous formation of small 
aggregates persisting up to tens of picoseconds at $x_{\text{\tiny{TBA}}} = 0.08$, but not at 
$x_{\text{\tiny{TBA}}} = 0.02$. At the same time, they also noted structural ordering of the surrounding water structure.
Several other simulation studies~\cite{fornili_pccp_2003, kerdcharoen_cpl_2003, paul_patey_jpcb_2006} revealed the 
self-association of TBA molecules in their binary mixture. Lee and Vegt~\cite{lee_vegt_tba_pot} proposed a modified 
force field for TBA in aqueous solution, in order to have a better approximation of the Kirkwood-Buff integrals 
and made detailed analysis of the structural aspects, again asserting the presence of TBA self-aggregates. 
Kezic and Perera~\cite{perera_jcp_2012} introduced a ``molecular emulsion'' picture, based on Teubner-Strey approach, 
to describe the microheterogeneity of water-TBA binary mixture. Recently, Gupta and Patey~\cite{gupta_patey_jcp_2012} 
have done an extensive simulation on water-TBA binary mixture using different force fields, by including up to 64000 
particles. Although their study indicated certain inconsistency in the force-field and system-size dependency of 
water-TBA solutions, the general picture of self-aggregation of TBA molecules remained undisputed. However, even using 
an effective fragment potential (EFP), wherein the parameters are derived from ab-initio calculations, Hands et 
al.~\cite{slipchenko_jpcb_2012} found that at low TBA concentrations, the structure of water is enhanced and water 
and TBA are not homogeneously mixed at the molecular level.

In the present work, we have used a system size of 3000 particles and the Lee-Vegt potential~\cite{lee_vegt_tba_pot} 
for describing the TBA. It has been earlier shown that these conditions sufficiently reproduce the general nature of 
this binary mixture.  We demonstrate that the anomalies can be understood in terms of a percolation transition of 
the solute molecules at relatively low concentration. At very low concentrations, the TBA molecules self-aggregate 
inducing microheterogeneity in the system. Beyond a critical concentration, the self-aggregated molecules start 
forming a spanning cluster culminating in a percolation transition. The physical and thermodynamic anomalies of the 
system are observed mostly in this concentration regime.

Our previous works have shown that such a percolation transition also occurs in the aqueous solutions of other 
amphiphilic solutes, like dimethyl sulfoxide (DMSO)~\cite{saikat_jpcb_2010_1, saikat_jpcb_2010_2} 
and ethanol (EtOH)~\cite{saikat_jpcb_2012}. We must emphasize that the present work is a continuation of this 
series of investigation with aim towards a unified understanding of aqueous binary mixtures of amphiphilic solutes.

The rest of this article is organized as follows. In Sec.~\ref{simulation} we report the details of simulation. 
In Sec.~\ref{tba-anomalies}, we have discussed the dependence of local structure and diffusion coefficient of 
TBA on composition of the binary mixture. 
In Sec.~\ref{tba-percolation}, we present the analysis of percolation, using the classical approach, as well as 
the fractal dimension approach. We also interpret the anomalies in terms of percolation. 
In Sec.~\ref{water-structure} we consider the structural changes in the water molecules alongside the percolation 
transition of TBA molecules. 
In Sec.~\ref{water-percolation} we have tried to locate the percolation threshold of water molecules, 
at relatively higher mole fraction of TBA. 
We present some snapshots of the simulation in Sec.~\ref{snapshots}. 
Finally, we have concluded the discussion in Sec.~\ref{conclusion}.

\section{Simulation Details \label{simulation}}

We have performed molecular dynamics (MD) simulation of the water-TBA binary mixture. All simulations have been 
done at 300 K temperature and 1 bar pressure. We have used the extended simple point charge model 
(SPC/E)~\cite{spce_water_01, spce_water_02} for water. We have treated the TBA molecules as united atoms, 
using the force field proposed by Lee and van der Vegt~\cite{lee_vegt_tba_pot}. To perform MD simulation, 
we have used GROMACS (version 4.5.5) which is highly scalable and efficient molecular simulation 
engine~\cite{gromacs_01, gromacs_02, gromacs_03, gromacs_04}.

We created solvent box containing TBA and water, performed energy minimization, and then equilibrated them 
for 2 ns, keeping the volume and temperature constant. After that, we again performed an equilibration at 
constant pressure and temperature for 2 ns before doing the production run for 20 ns at constant pressure 
and temperature. The cubic solvent box, in all cases, had 3000 particles with proper mole fraction ratio. 
The calculations of all the properties of the binary mixture were done using these trajectories.

Periodic boundary condition was applied in all the simulations. We have used Nose-Hoover 
thermostat~\cite{nose-hoover_01, nose-hoover_02} for temperature coupling and Parrinello-Rahman 
barostat~\cite{parinello-rahman} for pressure coupling. We have used a time step of 2 fs for all the 
simulations. We employed a grid system for neighbor searching while calculating the non-bonded interactions. 
Neighbor list generation was performed every 5 step. The cutoff radius for neighbor list and van der Waals 
interaction was 1.4 nm. To calculate the electrostatic interactions, we used particle mesh 
Ewald (PME)~\cite{pme_01, pme_02} with a grid spacing of 0.16 nm and an interpolation order of 4.

\section{Structural and dynamical characterization: Anomalies of the binary mixture \label{tba-anomalies}}

\begin{figure}
\subfigure[]{\includegraphics[width=0.48\textwidth]{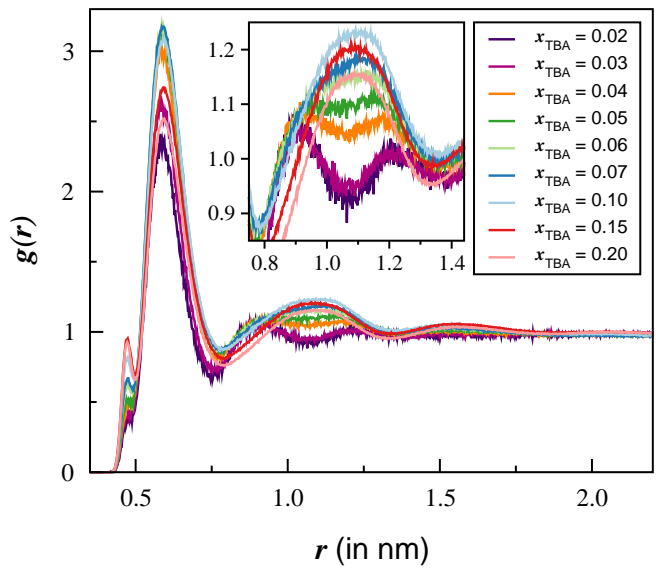}}
\vspace{10pt}
\subfigure[]{\includegraphics[width=0.48\textwidth]{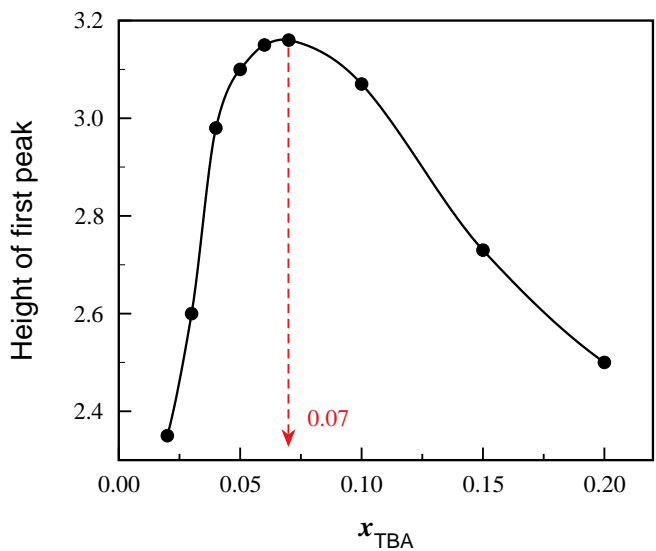}}
\caption{\label{tba-tba-gr}
(a) Radial distribution function (rdf) of the central C atom of TBA at various concentration of the aqueous 
binary mixture. The legend shows the mole fraction of TBA at various concentrations. 
Inset: Enlarged region of the second peak. 
(b) The height of the first peak of rdf is plotted against composition. There is a non-monotonic behavior in 
the peak height showing that there is enhanced ordering of TBA molecules at $x_{\text{\tiny{TBA}}} \approx 0.07$. 
Note the shift in position of the second peak as well.}
\end{figure}

Several experimental studies have revealed the composition dependent anomalies of water-TBA binary mixture. 
For example, partial molar volume~\cite{nakanishi_bcsj_1960}, 
excess heat capacity~\cite{desnoyers_jsolchem_1980}, 
ultrasonic absorption~\cite{kaatze_jpc_1992}, 
light scattering~\cite{fujiyama_jpc_1979, bender_pecora_jpc_1986, sorensen_jcp_1984}, 
etc show non-monotonic dependence on the composition. Most of these anomalies are observed at low 
concentrations of TBA $(x_{\text{\tiny{TBA}}} \approx 0.03 - 0.07)$. The radial distribution function (rdf) 
has always been an invaluable tool to get an insight of the structural properties of a system. Here we study 
the rdf of the central C atoms of TBA molecules at different concentrations of the binary mixture. A comparison 
of the rdf at different concentrations of the binary mixture is shown in Fig.~\ref{tba-tba-gr}a. We have focused 
on the low concentration regime of the binary mixture since the anomalies are mostly observed in this regime. 
We find that the height of the first peak of rdf has non-monotonic composition dependence, and is plotted in 
Fig.~\ref{tba-tba-gr}b. The peak height of the rdf increases up to $x_{\text{\tiny{TBA}}} \approx 0.07$ 
and then starts decreasing. The peak height gives a measure of the probability of finding the molecules at 
the given distance, i.e. it gives a measure of the relative concentration of the molecules. 
Hence, the relative concentration of TBA in the first hydration shell increases up to 
$x_{\text{\tiny{TBA}}} \approx 0.07$, followed by a decrease in the relative concentration. 
That means, as concentration of TBA is increased they initially fill up the first hydration shells 
(i.e. they form self-aggregates) before spanning out over the rest of the system.

The second peak of the rdf, shown in the inset of Fig.~\ref{tba-tba-gr}a, is also interesting. 
Initially, below $x_{\text{\tiny{TBA}}} \approx 0.05$, the second peak of the rdf appears at $0.9nm$. 
At $x_{\text{\tiny{TBA}}} \approx 0.05$, the second peak becomes flat and extended like a plateau. 
As we increase the concentration, the flat plateau-like region disappears and shows a distinct peak at $\sim 1.1nm$. 
This clearly indicates the enhancement of the density of TBA molecules in the second neighboring shell beyond 
$x_{\text{\tiny{TBA}}} \approx 0.05$. This is a unique behavior of aqueous TBA solution, and is appearing due 
to the aggregation of the TBA molecules. It is further discussed in Sec.~\ref{tba-percolation.percolation}.

\begin{figure}
\includegraphics[width=0.48\textwidth]{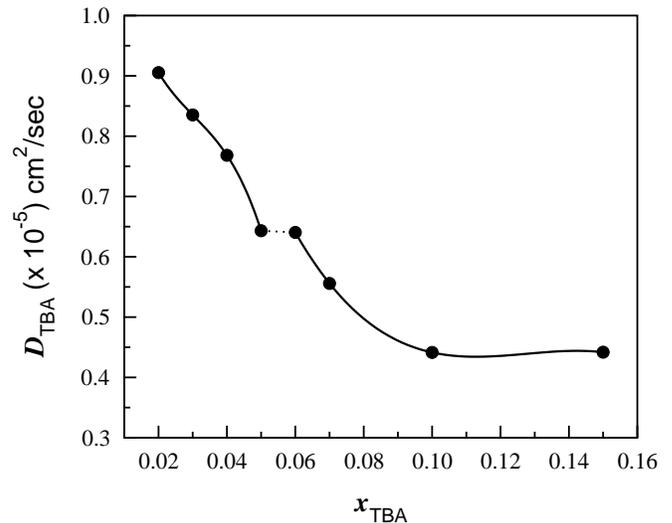}
\caption{\label{tba-diff}
Change of diffusion coefficient of TBA with composition. The diffusion coefficient at different concentrations 
is shown by solid circles, while the solid line is an aid to the eye. Note the abrupt change in slope of the 
diffusion coefficient before $x_{\text{\tiny{TBA}}} \approx 0.05$ and after $x_{\text{\tiny{TBA}}} \approx 0.06$, 
while the diffusion coefficient remains almost unchanged at $x_{\text{\tiny{TBA}}} \approx 0.05$ and $0.06$. }
\end{figure}

In order to give an overview of the dynamic anomalies, we evaluated the composition dependence of the 
self-diffusion coefficient of TBA and plotted the same in Fig.~\ref{tba-diff}. It reveals non-monotonic 
dependence on the concentration. The self-diffusion coefficient decreases as we increase the concentration 
of TBA, which is quite expected. However, there is a change in slope of the decrease at 
$x_{\text{\tiny{TBA}}} \approx 0.05 - 0.06$. This indicates an arrested motion of the TBA molecules in that 
concentration range, which will be further revealed in Sec.~\ref{tba-percolation.percolation}.

\begin{figure}
\includegraphics[width=0.48\textwidth]{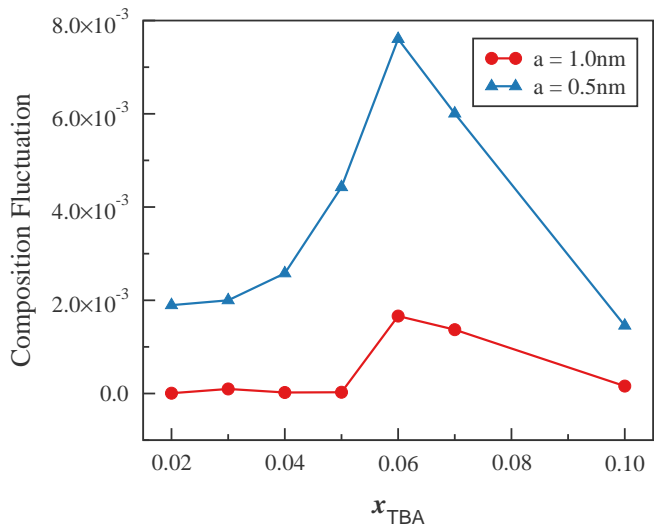}
\caption{\label{tba-compfluct}
Average local composition fluctuation of the binary mixtures at different concentrations. 
We consider a sphere of radius $a$ and calculate the local composition (TBA mole fraction) at every step. 
We note that there is a sharp deviation in the average local composition fluctuation, with a peak at
$x_{\text{\tiny{TBA}}} \approx 0.05$. The deviation becomes less prominent as we increase $a$, thereby 
showing that the fluctuation is a local phenomenon.}
\end{figure}

Next, we probed the local structure of the system. The origin of the non-monotonic composition-dependence 
of many physicothermal properties (like diffusion and viscosity) of binary mixtures can be understood from 
the composition fluctuation of the system, especially those at small length scales~\cite{murarka_jcp_2002, 
srini_arnab_jcp_2001}. Here, we studied the concentration dependence of average composition fluctuation of 
the system. In Fig.~\ref{tba-compfluct}, we show the concentration dependence of the local composition 
fluctuation $(\sigma_{x}^{2})$ of the system. We measure the mean square deviation of mole fraction of TBA 
in a sphere of a given radius $(a)$ at different concentrations of the binary mixture.
\begin{equation} \label{eq.tba-compfluct}
    \sigma_{x}^{2} = \left\langle(x_i - \bar{x})^2 \right\rangle
\end{equation}
where $x$ is the mole fraction of TBA in the mixture. Within a sphere of radius $0.5nm$, we see a sudden increase 
in the local composition fluctuation at a concentration, $x_{\text{\tiny{TBA}}} \approx 0.06$. However, the 
amplitude of the fluctuation is rather small for a larger spherical region. We would like to draw particular 
attention to the length-scale dependence of the composition fluctuation observed here. This is a matter of great 
importance because it may control the solvation of solutes in a non-trivial manner. While a small solute 
like methanol / ethanol is susceptible to such composition fluctuation, a much larger solute (like a protein) 
might see only an average composition. This is the importance of Fig.~\ref{tba-compfluct}.

\section{Percolation transition of TBA \label{tba-percolation}}
\subsection{Definition of clusters \label{tba-percolation.def-cluster}}
To understand the microheterogeneity of the system, we define self-aggregates of TBA as clusters. 
We follow the traditional method of defining connectivity in clusters of non-bonded systems through 
radial distribution functions (rdf). We measured the rdf of central C atoms of TBA. The first minimum 
of the rdf (Fig.~\ref{tba-tba-gr}a) gives us a measure of separation between the centers of two TBA molecules. 
Not only that the central C atom is very close to the center-of-mass of the molecule, but also the bulkiness 
of the t-butyl groups and the spherical electron density ensures that the central C-atoms should be within a 
particular distance if there is any hydrophobic association. We then defined the TBA clusters based on that 
criterion of mutual separation. We noted that if the central C-atoms are within a distance of $0.8nm$ then we 
can consider two TBA molecules to be associated together to form a cluster. A cluster is defined as a group 
of TBA molecules connected by this nearest neighbor distance.

\begin{figure}
\includegraphics[width=0.48\textwidth]{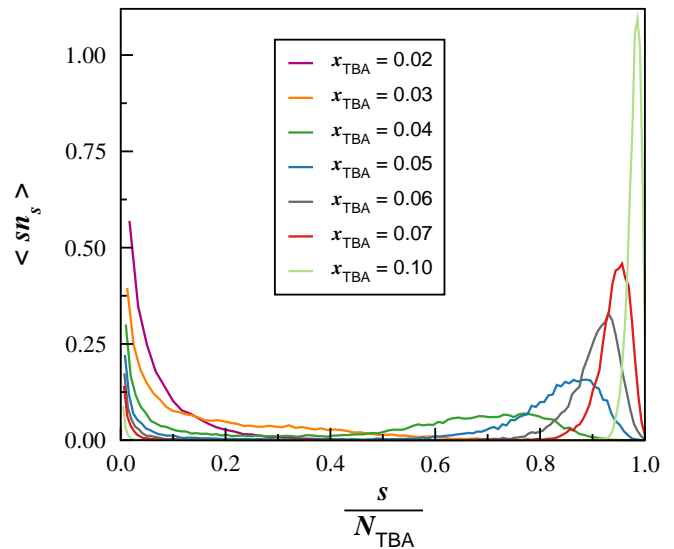}
\caption{\label{tba-cluster-size-distribution}
Plot of average number of TBA molecules in a cluster against the scaled cluster size. Here, $s$ is the size of 
the cluster, and $n_s$ is the number of $s$-sized cluster. At lower concentrations, the TBA molecules form smaller 
sized clusters. However, as the concentration increases, larger clusters start forming, and after a certain 
concentration, almost all molecules in the system are associated to form a single large cluster.}
\end{figure}

\subsection{Percolation transition \label{tba-percolation.percolation}}
In our calculations, we considered clusters of size $s$, with the fraction of $s$-sized clusters being $n_s$. 
We calculate the basic quantities involved in percolation~\cite{stauffer_percolation_review}. 
We plotted the first moment of $s$, i.e. $sn_{s}$ against $s$ for various concentration of the binary mixture, 
as shown in Fig.~\ref{tba-cluster-size-distribution}. At lower concentrations, we find that the system is 
microheterogeneous with several sized cluster prevailing in the system. But as we increase the concentration, 
the peak value of $\langle sn_{s} \rangle$ approaches 1, indicating that one continuous spanning cluster exists 
in the system. The bimodality~\cite{bagchi_gibbs_cpl_1983, bagchi_gibbs_prbcondmat_1981} of the cluster size 
distribution is an indicative feature of percolation transition.

\begin{figure}
\includegraphics[width=0.48\textwidth]{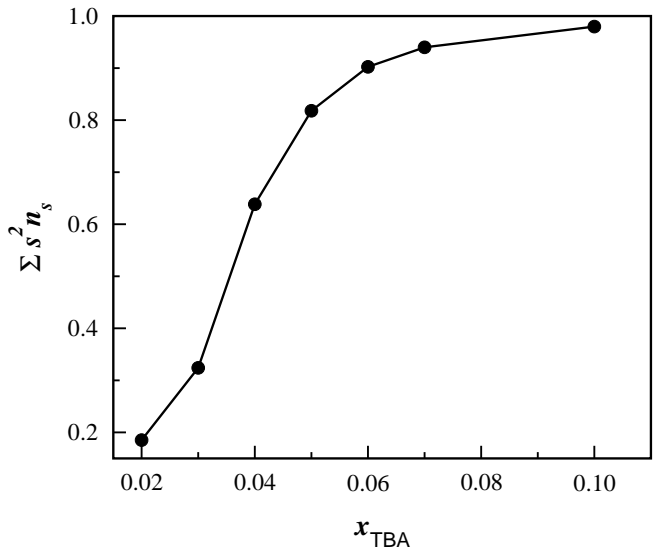}
\caption{\label{tba-percolation-op}
Plot of the order parameter in percolation transition of TBA. Here, $s$ is the size of the cluster, and 
$n_s$ is the number of $s$-sized cluster. The percolation threshold appears at $x_{\text{\tiny{TBA}}} \approx 0.05$.}
\end{figure}

In order to quantify the percolation transition, we calculate the order parameter for 
transition~\cite{stauffer_percolation_review}, $\sum{s^{2}n_s}$ , normalized by the number of TBA molecules present 
in the system. It is interesting to note here that this is a second moment term, which gives the susceptibility. 
In analogy with the standard approach of percolation theory, the different concentration of the binary mixture 
signifies the occupational probabilities. The summation is done over all possible cluster size, and the variation 
in order parameter over the concentration range is shown in Fig.~\ref{tba-percolation-op}.

We find here that TBA undergoes a percolation transition at $x_{\text{\tiny{TBA}}} \approx 0.05$. 
Surprisingly, most of the anomalies in the mixture are observed in the concentration range 
$x_{\text{\tiny{TBA}}} \approx 0.03 - 0.07$. These anomalies were attributed to structural transformations. 
It has been earlier suggested that clathrate-like alcohol hydrates are formed in dilute water-TBA 
solutions~\cite{granick_science_2002, fujiyama_jpc_1979, sorensen_jcp_1984, nishikawa_jpc_1987}. 
However, we do not find any such clathrate-like structures. Our results corroborates to the neutron diffraction 
study of Bowron, Finney and Soper~\cite{soper_finney_bowron_jpcb_1998}, who found similar self-aggregates of 
TBA molecules driven mainly by hydrophobic interactions of the tert-butyl groups. The presence of TBA 
aggregates leading to microheterogeneity in the mixture has earlier been shown in extensive simulations by 
Gupta and Patey~\cite{gupta_patey_jcp_2012}. However, as mentioned earlier, characterization of these 
clusters has been scarce~\cite{perera_jcp_2012}.

The percolation transition of the TBA clusters changes the nature of the binary mixture -- leading from 
microaggregates to a bi-continuous phase. The TBA molecules start forming aggregates in the concentration 
range of $x_{\text{\tiny{TBA}}} \approx 0.02 - 0.03$, and finally form a spanning cluster beyond
$x_{\text{\tiny{TBA}}} \approx 0.06$. The results presented here suggest that this microphase transition 
could be responsible for the anomalous behavior of water-TBA binary mixture. It is to be noted here that 
true ``percolation'' is discussed in terms of probabilistic description, but in this case, what we are 
observing is closer to, but not exactly, a bootstrap percolation. The crossover from microaggregates to a 
bi-continuous phase is not only facilitated by occupational probabilities but also the inherent nature of 
the TBA molecules to self-aggregate, driven by hydrophobic interaction. Without going into the details 
of the origin of hydrophobic interaction, which itself is a demanding discussion, we would rather consider 
the manifestation of this transition on composition-dependent anomalies. We will continue with the 
nomenclature of ``percolation transition'' to denote this clustering transition through the rest of this article.

We can now understand the origin of the representative anomalies that we have discussed earlier in 
Sec.~\ref{tba-anomalies}, viz. radial distribution function of central C atom of TBA, and diffusion 
co-efficient of TBA. At the percolation threshold, one should expect an extended but fluctuating network 
of TBA molecules (see Fig.~\ref{tba-cluster-size-fluctuation}), which is reflected in the increased height 
of the first peak, and a broad ``plateau-like'' second maximum. Beyond the percolation threshold, 
the extended network becomes stable and is reflected in the shift of the second peak. The first peak height 
decreases because the relative concentration in the first neighboring shell decreases since the TBA molecules 
are now forming a spanning cluster.

The self-diffusion coefficient of TBA is also directly related to the percolation transition. As shown in 
Fig.~\ref{tba-diff}, the rate of decrease in self-diffusion coefficient is distinctly different before and 
after percolation transition. Before percolation, the particles are free to move, whereas their motion is 
arrested when they form the spanning cluster beyond percolation threshold. Hence, it is quite obvious that 
the slope is much steeper before percolation, as compared to that after percolation. The indifference of the 
self-diffusion coefficient values at $x_{\text{\tiny{TBA}}} \approx 0.05$ and $0.06$ is clearly a direct 
consequence of the percolation threshold. This can be explained more easily if we look at the plot from 
the opposite direction. As we are decreasing the concentration of TBA, the self-diffusion coefficient is 
increasing. However, if we decrease the concentration from $x_{\text{\tiny{TBA}}} \approx 0.06$ to $0.05$, 
there is no decrease in mobility because all the TBA molecules are now involved in the spanning cluster. 
Hence, there is no change in the self-diffusion coefficient as well. Just below that, at 
$x_{\text{\tiny{TBA}}} \approx 0.04$, the spanning cluster breaks down to form smaller aggregates, 
their mobility increases and the self-diffusion coefficient takes a noticeable jump.

Our previous studies have shown that such percolation transition happens in case of other amphiphilic 
solutes as well. We have shown similar transitions in case of ethanol~\cite{saikat_jpcb_2012} and 
DMSO~\cite{saikat_jpcb_2010_1, saikat_jpcb_2010_2} at $x_{\text{\tiny{eth}}} \approx 0.10$ 
and $x_{\text{\tiny{DMSO}}} \approx 0.15$, respectively. In case of TBA, the percolation takes place at a much 
lower concentration (as compared to ethanol and DMSO) owing to the strong hydrophobicity of the t-butyl group. 
As mentioned earlier, such percolation transition is a general phenomenon common to this class of binary mixtures 
of amphiphilic solutes, but not appreciated sufficiently.

\subsection{Further evidence of percolation transition from fractal dimension \label{tba-percolation.fractal}}

We substantiate our findings by calculating the fractal dimension of the largest cluster at various concentrations 
of the binary mixture. It has been argued before that the \textit{largest cluster of a system is a fractal object above 
the percolation threshold and no objects with fractal dimension lower than 2.53 can be infinite in three-dimensional 
space}~\cite{oleinikova_fractal_dimension_1, oleinikova_fractal_dimension_2}. Hence, the true percolation threshold 
is located where the fractal dimension of the largest cluster in the system reaches the critical value of $2.53$.

\begin{figure}
\includegraphics[width=0.48\textwidth]{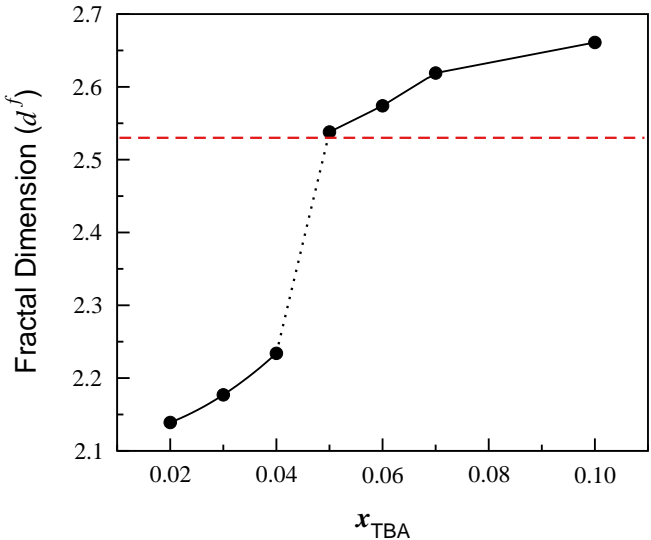}
\caption{\label{tba-percolation-df}
Plot of the fractal dimension, $d^f$ against mole fraction of TBA. The percolation threshold for a cluster 
in a three-dimensional system is located where the fractal dimension of the largest cluster reaches the 
critical value of 2.53. The critical value is shown by the red dotted line. $x_{\text{\tiny{TBA}}} \approx 0.05$ 
the $d^f$ just exceeds the critical value, indicating the percolation threshold.}
\end{figure}

The fractal dimension of the largest cluster has been evaluated using the sandbox method. The key idea is to 
measure from a chosen point of the largest cluster of the system, how many other points lie within a given radius. 
In effect, however, this gives the cumulative radial distribution function of the groups belonging to the largest 
cluster of the system. We denote this function as $m(r)$. In other words, $m(r)$ is the number of groups belonging 
to the largest cluster and located closer than the distance $r$ from the center of mass of the cluster. 
The fractal dimension, $d^f$ is then evaluated by fitting the following equation,
\begin{equation} \label{eq.fractal-dimension}
   m(r) \sim r^{d^f}
\end{equation}
We obtained the values of fractal dimension from the logarithmic plot (not shown here). 
In Fig.~\ref{tba-percolation-df}, we show the plot of the fractal dimensions, $d^f$ at various concentrations of 
TBA in the binary mixture. We find that the fractal dimension just reaches the critical value at 
$x_{\text{\tiny{TBA}}} \approx 0.05$. This proves that for the water-TBA binary mixture, the t-butyl groups 
form a percolating network. The percolation threshold as measured by standard approach and that from the 
fractal dimension are in exact agreement, and we can, therefore, safely conclude that the percolation threshold 
appears at $x_{\text{\tiny{TBA}}} \approx 0.05$.

\subsection{Divergence of mean square cluster size fluctuation \label{tba-percolation.clustersizefluct}}

\begin{figure}
\includegraphics[width=0.48\textwidth]{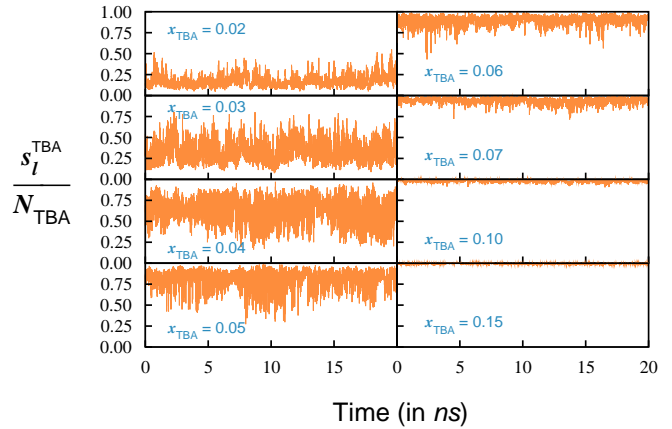}
\caption{\label{tba-cluster-size-fluctuation}
Size of the largest cluster of TBA, scaled by the number of TBA molecules, as a function of time. 
Note how the fluctuations increase with increasing concentration and then recedes after the percolation threshold.}
\end{figure}

\begin{figure}
\includegraphics[width=0.48\textwidth]{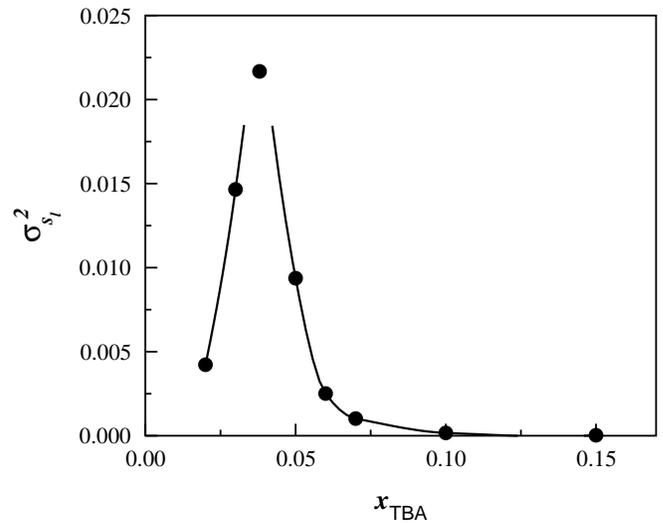}
\caption{\label{tba-cluster-size-fluctuation-op}
Standard deviation in the size of the largest cluster of TBA, as function of TBA mole fraction in the binary mixture. 
Note the divergent-like growth, reminiscent of a phase transition.}
\end{figure}

Percolation is a very weak phase transition, and critical phenomena are not generally observed in the 
thermodynamics of the system, but the structural changes are prominent. In case of TBA, we find large scale 
fluctuations in the size of the largest cluster. The size of the largest cluster, scaled by the number of 
TBA molecules, is plotted as a function of time in Fig.~\ref{tba-cluster-size-fluctuation}. With increasing 
concentration of TBA, the fluctuations in the largest cluster size increase up to the percolation threshold, 
above which it again start decreasing. Fluctuations are generally quantified by the standard deviation,
\begin{equation} \label{eq.cluster-fluctuation}
    \sigma_{s_l}^{2} = \left\langle (s_l - \langle s_l \rangle)^2 \right\rangle
\end{equation}
where $s_l$ is the size of the largest cluster. The standard deviation is plotted in 
Fig.~\ref{tba-cluster-size-fluctuation-op} as a function of TBA mole fraction. We find a divergence at the 
percolation threshold, indicative of a critical phenomenon. We anticipate that this phase transition is a 
universal phenomenon and should also be observed in case of other amphiphilic solutes in water.

\section{Structure of water in the mixture at low concentration: Observation of mild perturbation \label{water-structure}}

\begin{figure}
\includegraphics[width=0.48\textwidth]{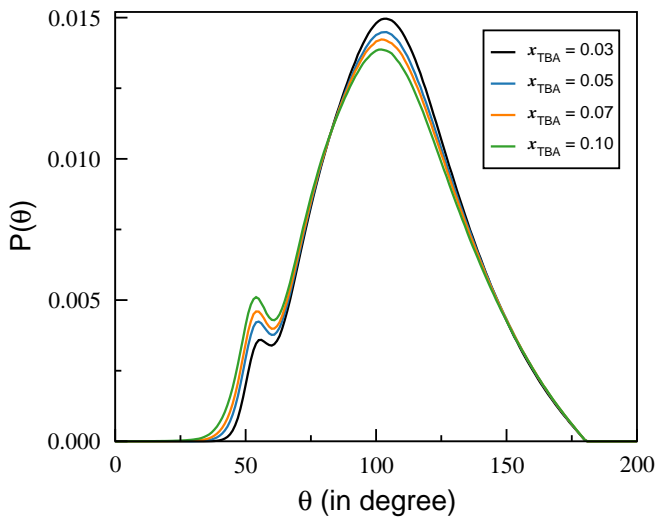}
\caption{\label{water-OOO-dist}
Distribution of O---O---O angles between water molecules at different concentrations of the binary mixture
(shown in legend). The fraction of water molecules with tetrahedral structure (peak height at $\sim104^\circ$)
is decreasing while the peak height of interstitial water molecules is increasing.}
\end{figure}

\begin{figure}
\vspace{20pt}
\includegraphics[width=0.48\textwidth]{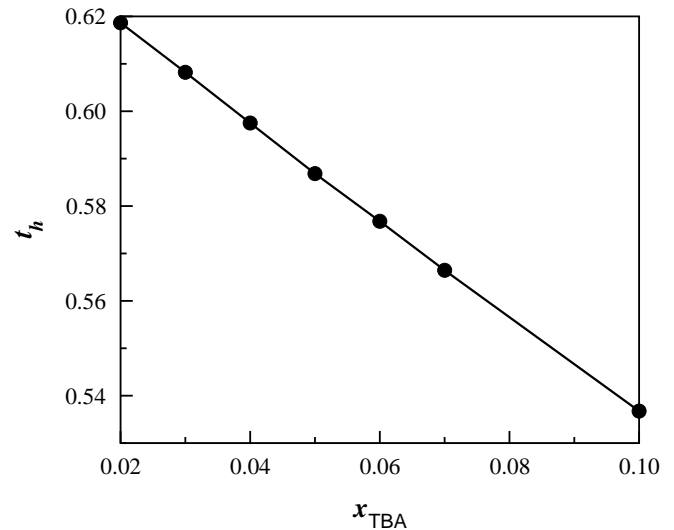}
\caption{\label{water-top}
The tetrahedral order parameter of water at different concentrations of the binary mixture. Note that TBA
is a structure breaker for water, and reduce the tetrahedrality. However, there is no significant non-monotonic
behavior in the concentration regime of percolation transition.}
\end{figure}

While we show that TBA molecules undergo a percolation transition in the water-TBA mixture, it is interesting 
to explore how the structure of water responds to the change. As shown by Gupta and Patey~\cite{gupta_patey_jcp_2012}, 
the radial distribution function of the oxygen atoms of water at $x_{\text{\tiny{TBA}}} \approx 0.03, 0.04, 0.05$ and 
$0.06$ does not show any significant difference in short length scales (even up to 8\AA{}). This indicates that the 
water-rich regions are not affected by the percolation transition. However, one might think that the orientation of 
the water molecules would respond to the strain caused by the spanning clusters of TBA. One useful way to capture 
the changes in water structure due to the hydrophobic association of the co-solvents is to calculate the tetrahedral 
order parameter $(t_h)$~\cite{stanley_tetrahedral_order_parameter}. It is defined as
\begin{equation}\label{eq.top}
 t_{h} = \frac{1}{n_{water}} \sum_{k}\left( 1 - \frac{3}{8}\sum_{i=1}^{3}\sum_{j=i+1}^{4}\left[\cos\psi_{ikj} - \frac{1}{3} \right]\right)
\end{equation}
where $\psi_{ikj}$ is the angle formed between the O atoms of the $k^{th}$ water molecule and the O atoms of 
the nearest neighbors, $i$ and $j$. In Fig.~\ref{water-OOO-dist}, we show the distribution of $\psi_{ikj}$, 
and in Fig.~\ref{water-top} we plot the $t_h$ values at various concentrations of the binary mixtures.

The O---O---O angle distribution (Fig.~\ref{water-OOO-dist}) clearly shows that number of water molecules with 
the tetrahedral orientation decrease with composition (peak height at $\sim 104^\circ$), while the interstitial 
water molecules increase (peak height at $\sim 60^\circ$). The change in orientation of the water molecules is 
quantitatively captured by the tetrahedral order parameter, which decreases monotonically (Fig.~\ref{water-top}) 
with increasing concentration of TBA in the binary mixture. Thus, TBA acts as a structure breaker for water, 
and reduces the tetrahedrality. Interestingly, the change in water structure is gradual and systematic in spite 
of the percolation of TBA molecules.

\begin{figure}
\includegraphics[width=0.48\textwidth]{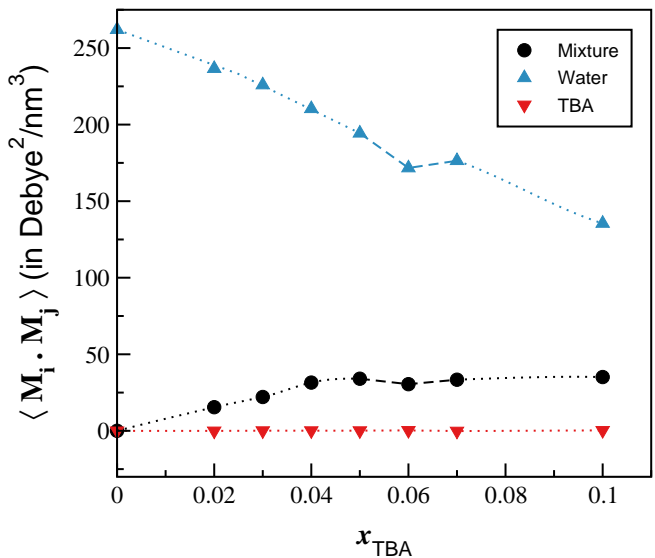}
\caption{\label{mu-cross-corr}
Cross correlation shown in terms of the dot product of the total dipole moment determined by Eq.~\ref{eq.dipole-moment-components} as a function of mole fraction.}
\end{figure}

It is also instructive to probe the dielectric constant of water, since it depends on the intermolecular 
correlation of the dipole moments, which in turn should be affected by the relative orientation of the molecules. 
Several theories have been established to highlight the importance of cross-correlations among the species in 
binary mixtures~\cite{amalendu_jcp_1989, amalendu_jcp_1991}. But they are difficult to address analytically 
for the molecules that are considered here. Nevertheless, we have studied the cross-correlation between the 
dipole moments of the components to understand the gross effect of change in orientation. The static dielectric 
constant, $\varepsilon(0)$ is given by,
\begin{equation}
    \varepsilon(0) = 1 + \frac{4\pi\left(\langle\mathbf{M}^2\rangle - \langle\mathbf{M}\rangle^2 \right)}{3k_{B}T \langle V \rangle}
\end{equation}
where $\mathbf{M}$ is the total dipole moment vector, $k_B$ is the Boltzmann constant, $T$ is the temperation and 
$V$ is the volume of the system. $\langle \ldots \rangle$ denotes time averaging. We are interested in the components of
$\mathbf{M}^2$, which in our case, are given by,
\begin{widetext}
\begin{equation}\label{eq.dipole-moment-components}
\mathbf{M}^2 = n_{\text{water}} |\boldsymbol\mu^w|^2 + n_{\text{\tiny{TBA}}} |\boldsymbol\mu^T|^2 + 2\left( \sum_{i \neq j}^{n_{\text{water}}} \boldsymbol\mu_{i}^{w} \cdot \boldsymbol\mu_{j}^{w} + \sum_{i \neq j}^{n_{\text{\tiny{TBA}}}} \boldsymbol\mu_{i}^{T} \cdot \boldsymbol\mu_{j}^{T} + \sum_{i \neq j}^{n_{\text{tot}}} \boldsymbol\mu_{i}^{w} \cdot \boldsymbol\mu_{j}^{T}\right)
\end{equation}
\end{widetext}
where $\boldsymbol\mu^{w}$ and $\boldsymbol\mu^{T}$ are dipole moments of water and TBA respectively, and 
$n_{\text{water}}$ and $n_{\text{\tiny{TBA}}}$ are respectively the number of molecules of water and TBA. 
The time averages of the correlation terms within the bracket are plotted in Fig.~\ref{mu-cross-corr} at different 
compositions of the binary mixture. We find a weak anomaly in the correlation of dipole moments among the water 
molecules, which is also reflected in the cross-correlation term. This is intriguing since it reflects the decrease 
in molecular level orientational correlation, and is a direct consequence of the strain in the solution imposed by 
the spanning of the TBA clusters. It signifies that there is a non-monotonic loss in orientational ordering among 
the dipoles of the water molecules, although the tetrahedral order parameter does not show such non-linearity.

\section{Percolation transition of water at higher concentration: Cooperative breakup of spanning water cluster \label{water-percolation}}

\begin{figure}
\includegraphics[width=0.48\textwidth]{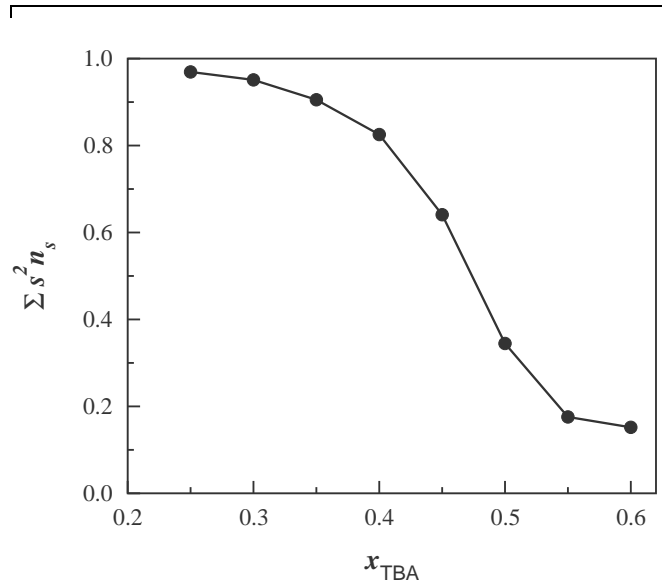}
\caption{\label{water-percolation-op}
Plot of the order parameter in percolation transition of water. Here, $s$ is the size of the water cluster, 
and $n_s$ is the number of $s$-sized cluster. The percolation threshold appears at $x_{\text{\tiny{TBA}}} \approx 0.45$.}
\end{figure}

\begin{figure}
\includegraphics[width=0.48\textwidth]{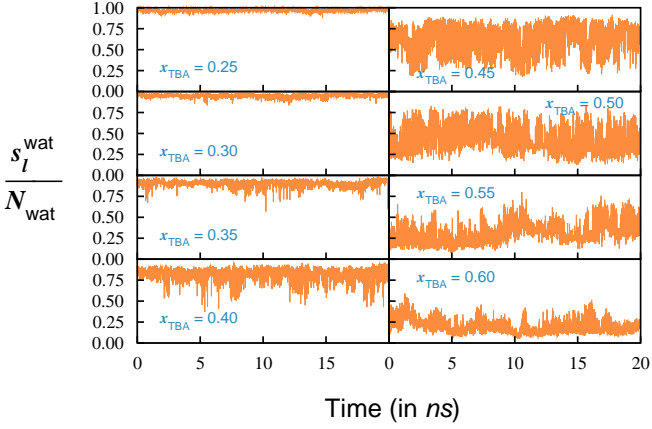}
\caption{\label{water-cluster-fluctuation}
Size of the largest cluster of water, scaled by the number of water molecules, as a function of time.}
\end{figure}

\begin{figure}
\includegraphics[width=0.48\textwidth]{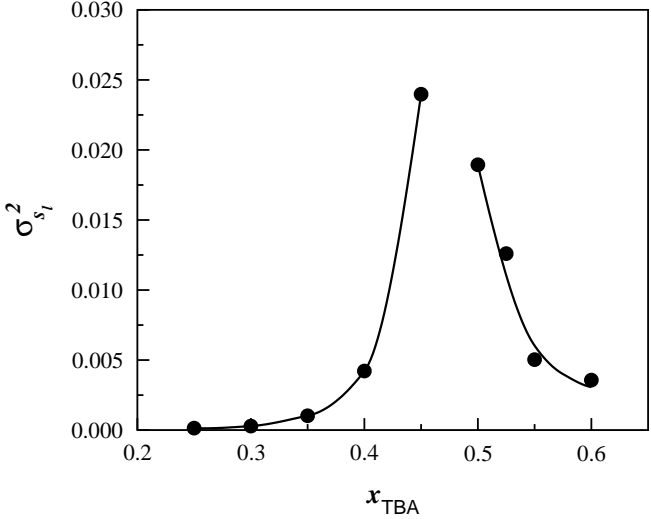}
\caption{\label{water-cluster-fluctuation-op}
Standard deviation in the size of the largest cluster of water, as function of TBA mole fraction in the binary mixture.}
\end{figure}

As the concentration of TBA is increased, another fascinating scenario unfolds, but this time involving 
water molecules themselves that forms a spanning cluster at low concentrations. It has earlier been pointed 
out that in case of other amphiphilic solutes water undergoes a percolation transition at higher 
concentration~\cite{susmita_jcp_2013}. This transition, however, is less elusive, as compared to the percolation 
of solute, and has been studied in great details for different mixtures. We define clusters using the criteria 
of H-bond. Different definitions have been used to estimate the H-bonds based on on the basis of various energy 
and structural criteria. Here we have adopted the geometric criterion proposed by 
Klein and co-workers~\cite{hbond_geometric_criterion}. If two water molecules are H-bonded, they are considered 
to be belonging to the same cluster. In Fig.~\ref{water-percolation-op} we plot the order parameter for percolation 
transition (as defined in Sec.~\ref{tba-percolation.percolation}), which shows that the percolation threshold 
appears at $x_{\text{\tiny{TBA}}} \approx 0.45$. This is also reflected in the fluctuation of the largest water 
cluster in the system. The fluctuations in the size of largest water cluster during the simulation time are 
shown in Fig.~\ref{water-cluster-fluctuation}. The standard deviation of the largest water cluster size, 
as defined in Sec.~\ref{tba-percolation.clustersizefluct}, at different mole fractions of TBA is shown in 
Fig.~\ref{water-cluster-fluctuation-op}. At the water percolation threshold, we find a significant divergence of 
the standard deviation -- very similar to that at the TBA percolation threshold.

\begin{figure}
\includegraphics[width=0.3\textwidth]{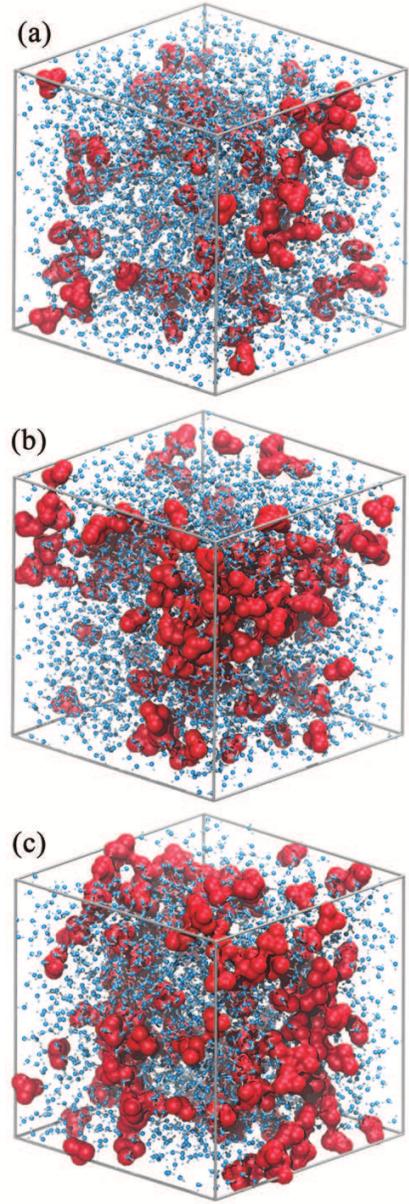}
\caption{\label{snap}
Snapshots of the simulation box at 3 different concentrations, (a) $x_{\text{\tiny{TBA}}} \approx 0.03$,
(b) $x_{\text{\tiny{TBA}}} \approx 0.05$ and (c) $x_{\text{\tiny{TBA}}} \approx 0.07$.
The surface of the aliphatic groups of the TBA molecules is shown in red, while the water molecules are shown in blue.
Note how the segregated ``islands'' at $x_{\text{\tiny{TBA}}} \approx 0.03$ forms a spanning cluster at the
percolation threshold, $x_{\text{\tiny{TBA}}} \approx 0.05$.}
\end{figure}

\section{Snapshots of the simulation \label{snapshots}}
In Fig.~\ref{snap}, we show some snapshots from our simulation to show the aggregation in this aqueous solution. 
We show the snapshots at three different concentrations -- one just before the onset of percolation transition of TBA 
$(x_{\text{\tiny{TBA}}} \approx 0.03)$, another at the percolation threshold $(x_{\text{\tiny{TBA}}} \approx 0.05)$ 
and the last one beyond the critical concentration of percolation $(x_{\text{\tiny{TBA}}} \approx 0.07)$. 
The TBA molecules are shown in red color in a surface representation, while the water molecules are shown in blue 
(using a different representation for clarity, since the water molecules will also form a continuous phase at this 
concentration). At low concentration, $x_{\text{\tiny{TBA}}} \approx 0.03$, before the onset of percolation transition, 
the co-solvents form ``islands''. We see that the TBA molecules are forming segregated clusters. 
At the critical concentration, $x_{\text{\tiny{TBA}}} \approx 0.05$ and beyond, we note that these ``islands'' 
have associated together to form spanning clusters, and there is an overall bi-continuous phase in the system. 
This microheterogeneity is of very low length and time scale so that they are not visible in the macroscopic phase.

\section{Conclusion \label{conclusion}}
In this work we have demonstrated the appearance of a percolation transition of the TBA clusters occurs 
at a composition range where the anomalies are most prominent. We have characterized the microheterogeneity 
using percolation theory. As mentioned earlier, we have found and reported in our previous works, similar 
percolation transition in aqueous binary mixtures of dimethyl sulfoxide and ethyl alcohol. These are all 
rather small amphiphilic solutes -- they have both hydrophobic and hydrophilic moieties in the same molecule. 
The hydrophobic groups form the core of the cluster while the hydrophilic tails remain in contact with water. 
The delicate balance in energetics helps the cluster to exist in a fluctuating microheterogeneous environment 
instead of segregating out. We believe that the percolation transition is a general phenomenon -- a phase transition 
in the cluster size distribution, which was hitherto unexplored. With a proper analytical description, 
one might be able to predict the structural features of these binary mixtures, which would be immensely helpful 
to tune the solvents towards greater utility. In a previous work~\cite{saikat_jcp_2013}, we studied the lifetimes 
of such regions and found that they are sensitive to the nature of the solute.

The complexity of intermolecular interactions of these solutes with water molecules and among themselves often 
precludes a detailed molecular theory of such anomalies~\cite{amalendu_jcp_1989,amalendu_jcp_1991}. 
A generalized explanation connecting the observed phenomena is still lacking. 
In progress of scientific knowledge, it is often useful to understand the 
phenomenon in representative subsets before reaching to a more general conclusion. 
Hence, understanding the anomalies of water-TBA solution is crucial and interesting.

\begin{acknowledgments}
It is a pleasure to thank Susmita Roy, Rikhia Ghosh and Dr. Rajib Biswas for help and discussions. 
This work was supported in parts by grants from BRNS (DAE) and DST, India. 
BB acknowledges support from JC Bose Fellowship (DST).
\end{acknowledgments}

%


\end{document}